\documentclass[aps,superscriptaddress,twocolumn,10pt,prx,floatfix,nofootinbib,sectionbib,sort&compress]{revtex4-2}

\input{glyphtounicode}
\usepackage[T1]{fontenc}
\usepackage[utf8]{inputenc}
\usepackage[english]{babel}
\usepackage{amsmath}  % needed for \tfrac, \bmatrix, etc.
\usepackage{amsfonts} % needed for bold Greek, Fraktur, and blackboard bold
\usepackage{amssymb}
\usepackage{graphicx} % needed for figures
\usepackage[pdftex,bookmarks=true,bookmarksopen,bookmarksnumbered,colorlinks,linkcolor=blue,citecolor=blue,urlcolor=blue]{hyperref}
\graphicspath{ {./} }

\usepackage{tabularx}
\usepackage{array}
\usepackage[table]{xcolor}
\usepackage{xcolor}
\usepackage{braket}
\usepackage{booktabs}
\usepackage{multirow}
\usepackage{caption}
\usepackage{hyperref}
\usepackage{dsfont} % for double lined 1 - identity symbol

\usepackage[resetlabels,labeled]{multibib}
\usepackage{comment}
\usepackage{siunitx}
\usepackage[shortlabels]{enumitem}

\newcommand{\aref}[1]{\hyperref[#1]{Appendix~\ref*{#1}}}
\DeclareCaptionLabelSeparator{line}{\hspace{2pt}\bf{|}\hspace{2pt}}

\begin{document}

\captionsetup[figure]{name={\bf{Fig.}},labelsep=line,justification=centerlast,font=small}
\renewcommand{\equationautorefname}{Eq.}
\renewcommand{\figureautorefname}{Fig.}
\renewcommand*{\sectionautorefname}{Sec.}
\renewcommand\thesection{\Alph{section}}

\title{Violating Bell's inequality in gate-defined quantum dots}

\author{Paul Steinacker}
\email{p.steinacker@unsw.edu.au}
\affiliation{School of Electrical Engineering and Telecommunications, University of New South Wales, Sydney, NSW 2052, Australia}
\author{Tuomo Tanttu}
\affiliation{School of Electrical Engineering and Telecommunications, University of New South Wales, Sydney, NSW 2052, Australia}
\affiliation{Diraq Pty. Ltd., Sydney, NSW, Australia}
\author{Wee Han Lim}
\affiliation{School of Electrical Engineering and Telecommunications, University of New South Wales, Sydney, NSW 2052, Australia}
\affiliation{Diraq Pty. Ltd., Sydney, NSW, Australia}
\author{Nard Dumoulin Stuyck}
\affiliation{School of Electrical Engineering and Telecommunications, University of New South Wales, Sydney, NSW 2052, Australia}
\affiliation{Diraq Pty. Ltd., Sydney, NSW, Australia}
\author{MengKe Feng}
\affiliation{School of Electrical Engineering and Telecommunications, University of New South Wales, Sydney, NSW 2052, Australia}
\affiliation{Diraq Pty. Ltd., Sydney, NSW, Australia}
\author{Santiago Serrano}
\affiliation{School of Electrical Engineering and Telecommunications, University of New South Wales, Sydney, NSW 2052, Australia}
\affiliation{Diraq Pty. Ltd., Sydney, NSW, Australia}
\author{Ensar Vahapoglu}
\affiliation{School of Electrical Engineering and Telecommunications, University of New South Wales, Sydney, NSW 2052, Australia}
\affiliation{Diraq Pty. Ltd., Sydney, NSW, Australia}
\author{Rocky Y. Su}
\affiliation{School of Electrical Engineering and Telecommunications, University of New South Wales, Sydney, NSW 2052, Australia}
\author{Jonathan Y. Huang}
\affiliation{School of Electrical Engineering and Telecommunications, University of New South Wales, Sydney, NSW 2052, Australia}
\author{Cameron Jones}
\affiliation{School of Electrical Engineering and Telecommunications, University of New South Wales, Sydney, NSW 2052, Australia}
\author{Kohei M. Itoh}
\affiliation{Department of Applied Physics and Physico-Informatics, Keio University, Yokohama 223-8522, Japan}
\author{Fay E. Hudson}
\affiliation{School of Electrical Engineering and Telecommunications, University of New South Wales, Sydney, NSW 2052, Australia}
\affiliation{Diraq Pty. Ltd., Sydney, NSW, Australia}
\author{Christopher C. Escott}
\affiliation{School of Electrical Engineering and Telecommunications, University of New South Wales, Sydney, NSW 2052, Australia}
\affiliation{Diraq Pty. Ltd., Sydney, NSW, Australia}
\author{Andrea Morello}
\affiliation{School of Electrical Engineering and Telecommunications, University of New South Wales, Sydney, NSW 2052, Australia}
\author{Andre Saraiva}
\affiliation{School of Electrical Engineering and Telecommunications, University of New South Wales, Sydney, NSW 2052, Australia}
\affiliation{Diraq Pty. Ltd., Sydney, NSW, Australia}
\author{Chih Hwan Yang}
\affiliation{School of Electrical Engineering and Telecommunications, University of New South Wales, Sydney, NSW 2052, Australia}
\affiliation{Diraq Pty. Ltd., Sydney, NSW, Australia}
\author{Andrew S. Dzurak}
\email{a.dzurak@unsw.edu.au}
\affiliation{School of Electrical Engineering and Telecommunications, University of New South Wales, Sydney, NSW 2052, Australia}
\affiliation{Diraq Pty. Ltd., Sydney, NSW, Australia}
\author{Arne Laucht} 
\email{a.laucht@unsw.edu.au}
\affiliation{School of Electrical Engineering and Telecommunications, University of New South Wales, Sydney, NSW 2052, Australia}
\affiliation{Diraq Pty. Ltd., Sydney, NSW, Australia}

\date{\today}

\begin{abstract}
%150 words limit for Nature Nanotechnology, no references
\textbf{
Superior computational power promised by quantum computers utilises the fundamental quantum mechanical principle of entanglement. 
However, achieving entanglement and verifying that the generated state does not follow the principle of local causality has proven difficult for spin qubits in gate-defined quantum dots, as it requires simultaneously high concurrence values and readout fidelities to break the classical bound imposed by Bell's inequality.
Here we employ heralded initialization and calibration via gate set tomography (GST), to reduce all relevant errors and push the fidelities of the full 2-qubit gate set above \SI{99}{\percent}, including state preparation and measurement (SPAM). We demonstrate a \SI{97.17}{\percent} Bell state fidelity without correcting for readout errors and violate Bell's inequality using direct parity readout with a Bell signal of S~=~\SI{2.731} close to the theoretical maximum of $2\sqrt{2}$. Our measurements exceed the classical limit even at elevated temperatures of \SI{1.1}{\kelvin} or entanglement lifetimes of $\SI{100}{\micro \second}$.
}

\end{abstract}
%Main text – up to 3,000 words, excluding abstract, Methods, references and figure legends.
\maketitle
Ever since Einstein and  Schr\"odinger discussed ``\textit{the characteristic trait of quantum mechanics}'' back in 1935~\cite{einstein_can_1935,schrodinger_discussion_1935}, scientists have been studying its mysterious properties, with Feynman proposing to harness it for quantum computing~\cite{feynman_simulating_1982,feynman_quantum_1986}. Relatively recently, in 2022, the Nobel Prize in Physics was awarded jointly to Alain Aspect, John F. Clauser, and Anton Zeilinger ``\textit{for experiments with entangled photons, establishing the violation of Bell inequalities and pioneering quantum information science}''~\cite{noauthor_celebrating_2022}. This is an appreciation of experimental implementations demonstrating non-locality of quantum mechanics with photons~\cite{clauser_proposed_1969,aspect_experimental_1982,aspect_experimental_1982-1,bouwmeester_experimental_1997,weihs_violation_1998} dating back to John Stewart Bell's suggested experiment in 1964~\cite{bell_einstein_1964}.

These early demonstrations did not consider all potential ``loopholes'' during experimental tests, which means that a local hidden variable theory could theoretically reproduce the gathered data~\cite{larsson_loopholes_2014}. 
So-called ``loophole-free'' Bell tests -- experiments closing all major loopholes simultaneously -- were demonstrated in 2015 and following years~\cite{hensen_loophole-free_2015,giustina_significant-loophole-free_2015,shalm_strong_2015,rosenfeld_event-ready_2017,li_test_2018} with NV centres in diamond and photons. 
In 2023 a loophole-free violation of Bell's inequality was demonstrated with superconducting qubits, where a \SI{30}{\meter} long cryogenic link was used in a remarkable effort to achieve spatial separation of the entangled qubits~\cite{storz_loophole-free_2023,storz_complete_2024}. 

Spin qubits in silicon are strong contenders for building a full-scale quantum computer due to their compatibility with semiconductor foundry processes~\cite{gonzalez-zalba_scaling_2021}. The first violation of Bell's inequality in silicon was demonstrated in an electron-nuclear donor spin system~\cite{dehollain_bells_2016}. 
While many demonstrations of Bell state tomography in gate-defined quantum dots have followed~\cite{watson_programmable_2018,leon_bell-state_2021,philips_universal_2022,xue_quantum_2022,noiri_fast_2022,mills_two-qubit_2022}, an experimental violation of Bell's inequality in gate-defined quantum dots is still missing as of yet. 

In this work, we violate Bell's inequality in gate-defined quantum dots close to the theoretical quantum correlation limit~\cite{cirelson_quantum_1980} and with $86\sigma$ confidence. We achieve this by operating electron spin qubits in silicon with state preparation and measurement (SPAM) and universal logic fidelities approaching the requirements for surface code error correction~\cite{raussendorf_fault-tolerant_2007,wang_surface_2011,fowler_towards_2012,madzik_precision_2022}. Even at elevated temperatures of $\SI{1.1}{\kelvin}$, we measure Bell signals above the classical limit $S=2$, with over $16\sigma$ confidence. Finally, we apply a dynamical decoupling sequence to store the generated entanglement for over $\SI{100}{\micro \second}$. 

Full-scale fault-tolerant quantum computing processors require quantum logic operations across the entire chip with errors below the quantum error correction threshold to harness their full capabilities. An experimental violation of Bell's inequality is a major milestone for every qubit platform as it provides a meaningful performance benchmark requiring simultaneously high fidelities in state preparation, manipulation and measurement in a single quantum information processor.

\section*{Device and two-qubit operation}

\begin{figure*}[ht!]
    %\hspace{-0.6cm}
    \includegraphics[width=1\textwidth,angle = 0]{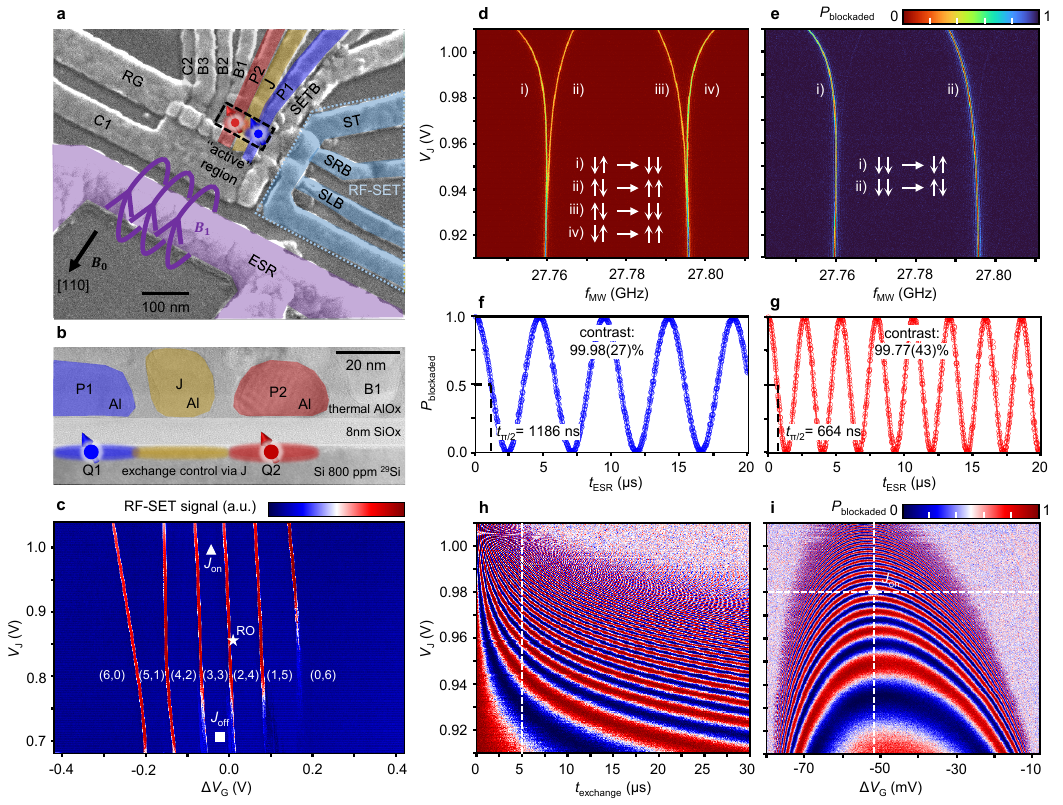}
    \caption{\textbf{Device and basic operation.}
    \textbf{a}, Scanning electron micrograph of a device nominally identical to that used in this work. Active gate electrodes and the microwave antenna are highlighted with colours. An external d.c.\;magnetic field $B_0$ and the antenna-generated a.c.\;magnetic field $B_1$ are indicated with arrows. The system operates at $T = \SI{0.1}{\kelvin}$, unless otherwise specified.
    \textbf{b}, Transmission electron micrograph of the ``active'' region with schematics indicating the quantum dot and electron spin qubit formation at the Si/SiOx interface including exchange control.
    \textbf{c}, Charge stability diagram as a function of P1, P2 voltage detuning $\Delta V_\mathrm{G} = -\Delta V_\mathrm{P1} = \Delta V_\mathrm{P2}$ and the J gate voltage $V_\mathrm{J}$, showing six loaded electrons across the double-dot system. The operation points for readout (RO), single-qubit operation ($J_{\rm off}$) and two-qubit operation ($J_{\rm on}$) are labelled as star (\scalebox{1.1}{$\star$}), square (\scalebox{0.5}{$\blacksquare$}), and triangle (\scalebox{0.75}{$\blacktriangle$}), respectively.
    \textbf{d, e}, Probability of detecting a blockaded state, $P_\mathrm{blockade}$, after a microwave burst of fixed power and duration at different J gate voltages $V_\mathrm{J}$ when preparing a mixed odd state $\frac{1}{\sqrt{2}}(\ket{\downarrow\uparrow}+\ket{\uparrow\downarrow})$ (\textbf{d}) and a pure state $\ket{\downarrow\downarrow}$ (\textbf{e}). The power and duration of the microwave burst are roughly calibrated to a single-qubit $\pi$-rotation. The following experiments are conducted with $\ket{\downarrow\downarrow}$ initialization, unless otherwise specified.
    \textbf{f, g}, Q1 and Q2 single-qubit Rabi oscillations at $V_\mathrm{J}=\SI{0.71}{\volt}$ as a function of pulse time $t_\mathrm{ESR}$, respectively.
    \textbf{h}, Decoupled controlled phase (DCZ) oscillations as a function of exchange time $t_\mathrm{exchange}$ and $V_\mathrm{J}$.
    \textbf{i}, $\mathrm{DCZ}$ exchange oscillation fingerprint for fixed exchange time $t_\mathrm{exchange} = \SI{5}{\micro \second}$ as a function of $\Delta V_\mathrm{G}$ and $V_\mathrm{J}$. Readout probability is unscaled in all data. Error bars represent the \SI{95}{\percent} confidence level.
    }
    \label{fig:main_fig_1}
\end{figure*}

\noindent We operate the silicon-metal-oxide-semiconductor (SiMOS) device (Fig.~\ref{fig:main_fig_1}a,b) in a double quantum dot with three electrons in each dot isolated from the reservoir (Fig.~\ref{fig:main_fig_1}c). Under influence of an external d.c.\;magnetic field the unpaired electron spin is an effective two-level system operated as a qubit~\cite{veldhorst_spin-orbit_2015,leon_coherent_2020}. The quantum dots are electrostatically confined by a multi-layer aluminium gate-stack~\cite{angus_gate-defined_2007} fabricated on top of an isotopically enriched ${}^{28}$Si substrate with $\SI{800}{ppm}$ residual ${}^{29}$Si~\cite{itoh_isotope_2014}. The device is biased such that the quantum dots are separated by approximately $\SI{60}{\nano\meter}$ occupying around $\SI{80}{\nano\meter^2}$ underneath the plunger gates (P1, P2) at the Si/SiO$_2$ interface. The exchange gate (J) in between gives control over the inter-dot separation and two-qubit exchange~\cite{loss_quantum_1998,petta_coherent_2005,cifuentes_bounds_2024} at an exponential rate of $\SI{24}{dec~V^{-1}}$ (Fig.~\ref{fig:main_fig_1}d). For single-shot charge readout, we integrate for $t_\mathrm{RO} = \SI{100}{\micro \second}$ the radiofrequency single-electron transistor (RF-SET)~\cite{angus_silicon_2008} signal at $\SI{165}{\mega \hertz}$. An on-chip antenna delivers the a.c.\;magnetic field $B_1$ to drive the electron spin state transitions.

A mixed odd state is initialized with a $t_\mathrm{init} = \SI{2}{\micro \second}$ ramp at $V_\mathrm{J} = \SI{0.86}{\volt}$ across the (2,4) to (3,3) inter-dot charge transition (Fig.~\ref{fig:main_fig_1}c,d). A pure $\ket{\downarrow \downarrow}$ state is obtained by heralding this initialization processes~\cite{huang_high-fidelity_2024} at an increased $V_\mathrm{J} = \SI{0.96}{\volt}$ (Fig.\ref{fig:main_fig_1}e).

Single-qubit gates are performed at $V_\mathrm{J} = \SI{0.71}{\volt}$ ($J_\mathrm{off}$) to minimize residual exchange interaction, $< \SI{10}{\kilo \hertz}$, between the two electron spins. Figures~\ref{fig:main_fig_1}f,g show coherent Rabi oscillations of both qubits. Our two-qubit gates are implemented as decoupled controlled phase gates (DCZ)~\cite{watson_programmable_2018,xue_quantum_2022} that are performed at $V_\mathrm{J} = \SI{0.98}{\volt}$ ($J_\mathrm{on}$). Figures~\ref{fig:main_fig_1}h,i show exchange oscillations and the exchange fingerprint map at $t_\mathrm{exchange} = \SI{5}{\micro \second}$.

We read out (RO) the spin state parity based on Pauli spin blockade (PSB)~\cite{seedhouse_pauli_2021}. Charge movement near the inter-dot charge transition from (3,3) to (2,4) is blockaded when both unpaired spins are parallel.

\section*{Two-qubit benchmarking}
\begin{figure*}[htb!]
    %\hspace{-0.6cm}
    \includegraphics[width=1\textwidth,angle = 0]{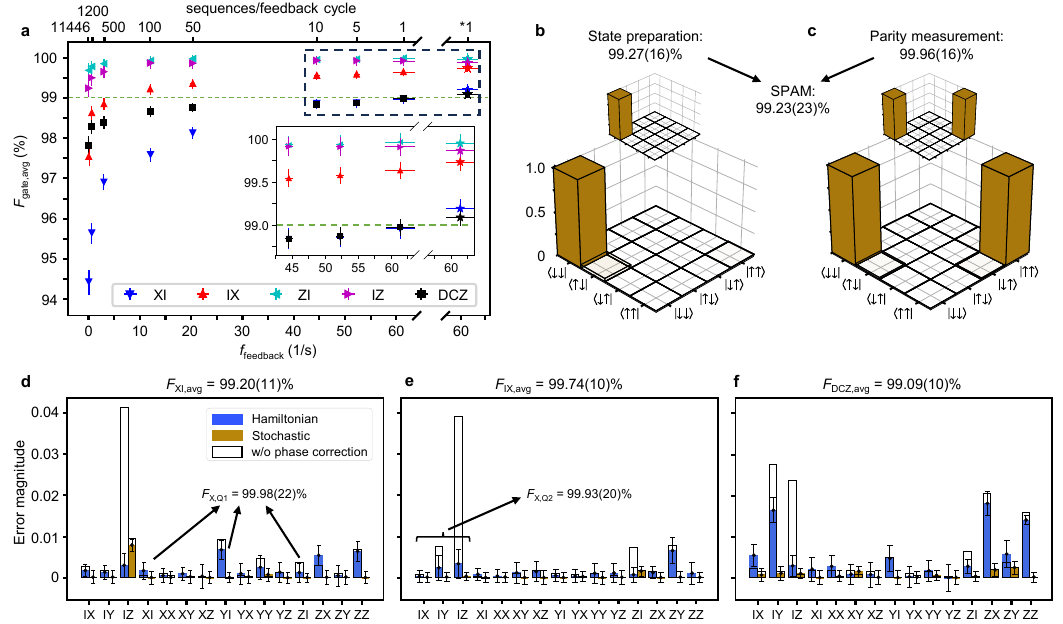}
    \caption{\textbf{Two-qubit benchmarking using GST.} 
    \textbf{a}, Gate infidelity as a function of the Larmor frequency feedback rate and number of GST sequences per feedback cycle. The star (\scalebox{1.1}{$\star$}) indicates additional phase corrections based on the previous GST results. The green dashed line indicates the commonly considered \SI{99}{\percent} threshold. The inset is a zoom-in of the black-dashed box.
    \textbf{b,c}, State preparation and measurement (SPAM) matrix, respectively. The insets show the respective theory matrix.
    \textbf{d-f}, Error magnitude of error components for the $\mathrm{XI}$, $\mathrm{IX}$, and $\mathrm{DCZ}$ gates from GST with (coloured bars) and without (uncoloured bars) additional phase correction. The average gate fidelity is given above each plot for the phase corrected GST measurement. The on-target X gate fidelity $F_\mathrm{X,Qi}$ can be calculated from the relevant error components. Hamiltonian errors contribute to the fidelity in second order, while stochastic errors contribute in first order.
    Error bars represent the \SI{95}{\percent} confidence level.
    }
    \label{fig:main_fig_2}
\end{figure*}

\noindent In Fig.~\ref{fig:main_fig_2}a we plot the full 2-qubit gate set's average gate fidelity benchmarked by gate set tomography (GST)~\cite{blume-kohout_demonstration_2017,nielsen_gate_2021} as a function of the Larmor frequency feedback rate $f_\mathrm{feedback}$. Faster feedback rates allow us to achieve significant improvements in the two-qubit XI and IX gates from $F_\mathrm{XI,avg} = \SI{94.42\pm 0.31}{\percent}$ and $F_\mathrm{IX,avg} = \SI{97.54\pm 0.23}{\percent}$ up to $F_\mathrm{XI,avg} = \SI{98.96\pm 0.12}{\percent}$ and $F_\mathrm{IX,avg} = \SI{99.64\pm 0.10}{\percent}$ by reducing the stochastic IZ and ZI error components attributed to phase loss~\cite{tanttu_consistency_2023}. Furthermore, the entangling DCZ gate is improved from $F_\mathrm{DCZ,avg} = \SI{97.82\pm 0.24}{\percent}$ to $F_\mathrm{DCZ,avg} = \SI{98.98\pm 0.10}{\percent}$. The error components with the most significant reduction in infidelity contribution are the stochastic XI, YI, XZ, ZX, and YZ (see Extended Data Fig.~\ref{fig:LarmorFeedback_GST_breakdown}). Additionally, we use the GST results to apply an informed phase correction calibrating for residual Larmor frequency mismatch (data point \scalebox{1.1}{$\star$} in Fig.~\ref{fig:main_fig_2}a).

Using these corrections, we can push all average gate fidelities above the commonly targeted threshold of $\SI{99}{\percent}$, including the state preparation and measurement (SPAM) fidelity (Fig.~\ref{fig:main_fig_2}b,c): $F_\mathrm{XI,avg} = \SI{99.20 \pm 0.11}{\percent}$, $F_\mathrm{IX,avg} = \SI{99.74 \pm 0.10}{\percent}$, $F_\mathrm{ZI,avg} = \SI{99.96 \pm 0.11}{\percent}$, $F_\mathrm{IZ,avg} = \SI{99.87 \pm 0.11}{\percent}$, $F_\mathrm{DCZ,avg} = \SI{99.09 \pm 0.10}{\percent}$, and $F_\mathrm{SPAM} = \SI{99.23 \pm 0.23}{\percent}$. 
Figures~\ref{fig:main_fig_2}d-f compare the error magnitude of the XI, IX, and DCZ gate with and without the additional phase correction informed by GST. We calibrated the correction to minimize the Hamiltonian IZ and ZI phase error components, without affecting other error components. To emphasize the distinction from the average gate fidelities in the two-qubit context, we also calculate the single-qubit gate fidelities, which are $F_\mathrm{X,Q1} = \SI{99.98 \pm 0.22}{\percent}$ and $F_\mathrm{X,Q2} = \SI{99.93 \pm 0.20}{\percent}$.

\section*{Bell test}
\begin{figure*}[ht!]
    %\hspace{-0.6cm}
    \includegraphics[width=1\textwidth,angle = 0]{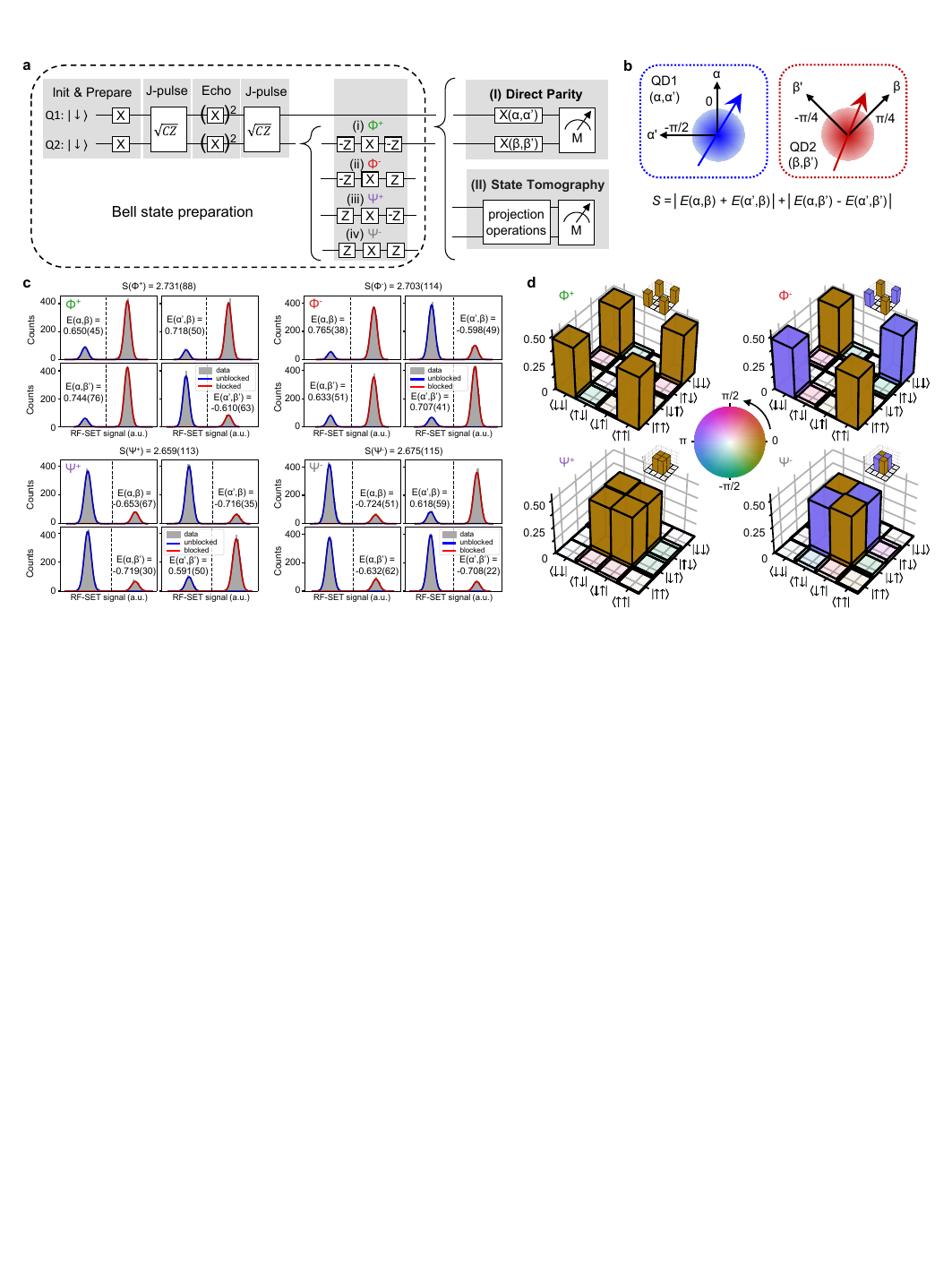}
    \caption{\textbf{Bell test.} 
    \textbf{a}, Protocol for conducting the Bell test in a gate-defined double-dot electron spin system. After preparation of a maximally entangled Bell state (i-iv), the quantum correlation is measured via (I) a direct parity measurement after rotation of each qubit to obtain the desired combination of projection axes in two bases, rotated by $\pi/4$, or (II) quantum state tomography.   
    \textbf{b}, Schematic of the two projection basis $(\alpha, \alpha')$ and $(\beta, \beta')$ of the electron spin qubit in quantum dot 1 and 2, respectively.  
    \textbf{c}, Histograms of RF-SET readout signal for all four Bell states in all possible combinations of axis projections at $T = \SI{0.1}{\kelvin}$. The data is fitted with a bimodal Gaussian distribution. The intersect of the two Gaussian curves is indicated by a dashed line defining the threshold for distinguishing odd (unblockaded) and even (blockaded) parity.
    \textbf{d}, Quantum state tomography results for all four Bell states at $T = \SI{0.1}{\kelvin}$. No corrections have been applied to compensate for initialization and readout errors. Insets indicate the theoretical density matrix of each Bell state.
    Error bars represent the \SI{95}{\percent} confidence level.
    }
    \label{fig:main_fig_3}
\end{figure*}

\noindent Figure~\ref{fig:main_fig_3}a shows the Bell experiment protocol; starting from a $\ket{\downarrow\downarrow}$ state, we prepare one of the four Bell states\: (i) $\Phi^+$, (ii) $\Phi^-$, (iii) $\Psi^+$, and (iv) $\Psi^-$, followed by measurement via either a (I) rotated basis parity readout or (II) quantum state tomography. We use the latter to confirm generation of the four maximally entangled Bell states with fidelities of $F_\mathrm{\Phi^{+}} = \SI{97.17 \pm 0.31}{\percent}$, $F_\mathrm{\Phi^{-}} = \SI{96.94 \pm 0.26}{\percent}$, $F_\mathrm{\Psi^{+}} = \SI{96.50 \pm 0.38}{\percent}$, and $F_\mathrm{\Psi^{-}} = \SI{96.47 \pm 0.31}{\percent}$, uncorrected for SPAM errors and at base temperature $T = \SI{0.1}{\kelvin}$ (Fig.~\ref{fig:main_fig_3}d). 

Bell's theorem~\cite{bell_einstein_1964} provides a means to experimentally verify that local hidden-variables do not play a role in quantum mechanics. This is done through the violation of the Clauser-Horne-Shimony-Holt (CHSH) inequality~\cite{clauser_proposed_1969}, and requires measurement of the quantum correlations of the two-qubit spin pair along all combinations of measurement bases $\alpha=0$, $\alpha'=-\pi/2$, $\beta=\pi/4$, and $\beta'=-\pi/4$ (Fig.~\ref{fig:main_fig_3}b). With parity readout, the quantum correlation in each of the four bases $(\alpha, \beta)$, $(\alpha', \beta)$, $(\alpha, \beta')$, and $(\alpha', \beta')$ becomes 
\begin{align}
\begin{split}
    E & = \dfrac{N_{\uparrow\uparrow}-N_{\uparrow\downarrow}-N_{\downarrow\uparrow}+N_{\downarrow\downarrow}}{N_{\uparrow\uparrow}+N_{\uparrow\downarrow}+N_{\downarrow\uparrow}+N_{\downarrow\downarrow}} %\\
    %& = \dfrac{N_{\uparrow\uparrow}-N_{\uparrow\downarrow}-N_{\downarrow\uparrow}+N_{\downarrow\downarrow}}{N_{\text{total}}} \\
      = \dfrac{N_{\text{even}}-N_{\text{odd}}}{N_{\text{total}}} \\
    & = P_{\text{even}} - P_{\text{odd}} %\\
      = 2P_{\text{even}} - 1,
\end{split}
\end{align}
with the number of even, odd, and total number of readout events $N_{\mathrm{even}} = N_{\mathrm{\uparrow\uparrow}} + N_{\mathrm{\downarrow\downarrow}}$, $N_{\mathrm{odd}} = N_{\mathrm{\uparrow\downarrow}} + N_{\mathrm{\downarrow\uparrow}}$, and $N_{\mathrm{total}} = N_{\mathrm{even}} + N_{\mathrm{odd}}$, respectively. Additionally, we use that the readout signal is either even or odd ($P_\mathrm{odd} = 1 - P_\mathrm{even}$). With that, the CHSH inequality becomes
\begin{align}\label{eq:CHSH}
\begin{split}
     S & = E(\alpha,\beta) - E(\alpha,\beta') + E(\alpha',\beta) + E(\alpha',\beta') \\
     & = 2(P_{\text{even}}^{\alpha,\beta} - P_{\text{even}}^{\alpha,\beta'} + P_{\text{even}}^{\alpha',\beta} + P_{\text{even}}^{\alpha',\beta'} - 1) \geq 2,
\end{split}
\end{align}
and proves Bell's theorem, if a Bell signal $S > 2$ is measured.   

Hence, we can utilize parity measurements in these bases to obtain a direct insight into the correlation of quantum entanglement of the spin system. Figure~\ref{fig:main_fig_3}c shows histograms of the RF-SET readout signal for all measurement basis combinations $(\alpha, \beta)$, $(\alpha', \beta)$, $(\alpha, \beta')$, and $(\alpha', \beta')$ for the four maximally entangled Bell states. Bimodal Gaussian fits allow us to extract the charge readout fidelity~\cite{serrano_improved_2024} and the threshold used to best distinguish between even and odd parity. The measured Bell signals $S_\mathrm{\Phi^{+}} = \SI{2.731 \pm 0.088}{}$, $S_\mathrm{\Phi^{-}} = \SI{2.703 \pm 0.114}{}$, $S_\mathrm{\Psi^{+}} = \SI{2.659 \pm 0.113}{}$, and $S_\mathrm{\Psi^{-}} = \SI{2.675 \pm 0.115}{}$ are up to more than $16\sigma$ above the classical limit $S=2$ from Bell's theorem. 

An alternative way to calculate the even parity probability $P_\mathrm{even}$ is by analytically transforming the density matrices measured via quantum state tomography (Fig.~\ref{fig:main_fig_3}d) into the rotated bases
\begin{align}
\begin{split}
    P_{\text{even}}^{\alpha\beta} & = P_{\downarrow\downarrow}^{\alpha\beta} + P_{\uparrow\uparrow}^{\alpha\beta} \\
    & = \left[R(\alpha) \otimes R(\beta) \times \rho \times \overline{R(\alpha)\otimes R(\beta)}\right]_{00} \\
    & \,+ \left[R(\alpha) \otimes R(\beta) \times \rho \times \overline{R(\alpha)\otimes R(\beta)}\right]_{33},
\end{split}
\end{align}
with the X gate rotation matrix
\begin{align}
    R_\mathrm{X}(\varphi) = 
    \begin{bmatrix}
        \cos{\varphi/2} & -i\sin{\varphi/2}\\
         -i\sin{\varphi/2}& \cos{\varphi/2}
    \end{bmatrix}.
\end{align}
By respective combinations of $\alpha=0$, $\alpha'=-\pi/2$, $\beta=\pi/4$, and $\beta'=-\pi/4$ we can calculate the Bell signal $S$ according to Eq.~(\ref{eq:CHSH}). At base temperature $T = \SI{0.1}{\kelvin}$ we achieve $S_\mathrm{\Phi^{+}} = \SI{2.721 \pm 0.030}{}$, $S_\mathrm{\Phi^{-}} = \SI{2.711 \pm 0.028}{}$, $S_\mathrm{\Psi^{+}} = \SI{2.703 \pm 0.032}{}$, and $S_\mathrm{\Psi^{-}} = \SI{2.693 \pm 0.016}{}$ with up to more than $86\sigma$ above the classical limit $S=2$ from Bell's theorem. Comparing both methods we get matching and consistently high Bell signals.

\section*{Bell test -- temperature dependence}

\begin{figure}[ht!]
    %\hspace{-0.6cm}
    \includegraphics[width=0.5\textwidth,angle = 0]{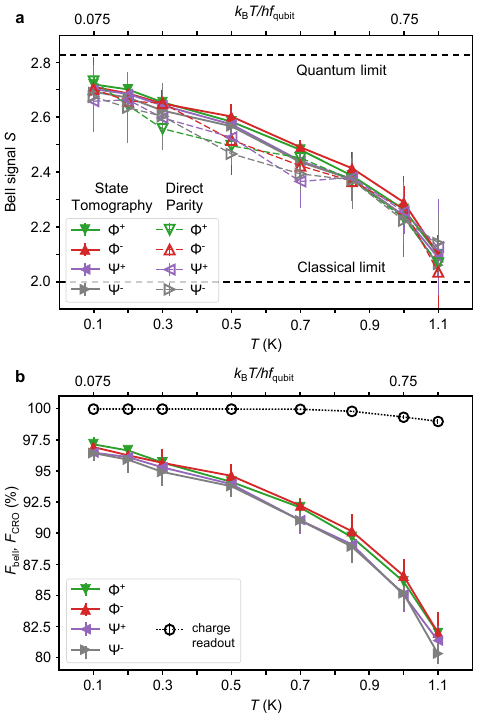}
    \caption{\textbf{Bell test -- temperature dependence.} Error bars represent the \SI{95}{\percent} confidence level.
    \textbf{a}, Bell signal $S$ as a function of temperature $T$ for all four maximally entangled Bell states measured by the direct parity measurement (open symbols) and quantum state tomography (filled symbols).
    \textbf{b}, Bell state ($F_{\rm Bell}$) and charge readout ($F_{\rm CRO}$) fidelities as a function of temperature $T$ for all four maximally entangled Bell states obtained from quantum state tomography and RF-SET signal histograms, respectively.
    Error bars represent the \SI{95}{\percent} confidence level.
    }
    \label{fig:main_fig_4}
\end{figure}

\noindent We extend the violation of Bell's inequality to operation temperatures of up to $\SI{1.1}{\kelvin}$ in Fig.~\ref{fig:main_fig_4}a. We maintain Bell signals of $S_\mathrm{\Phi^{+}} = \SI{2.101 \pm 0.064}{}$, $S_\mathrm{\Phi^{-}} = \SI{2.100 \pm 0.071}{}$, $S_\mathrm{\Psi^{+}} = \SI{2.088 \pm 0.053}{}$, and $S_\mathrm{\Psi^{-}} = \SI{2.061 \pm 0.039}{}$ measured via state tomography (filled symbols), and $S_\mathrm{\Phi^{+}} = \SI{2.068 \pm 0.154}{}$, $S_\mathrm{\Phi^{-}} = \SI{2.036 \pm 0.181}{}$, $S_\mathrm{\Psi^{+}} = \SI{2.127 \pm 0.176}{}$, and $S_\mathrm{\Psi^{-}} = \SI{2.142 \pm 0.109}{}$ measured via basis rotation, i.e., direct parity measurement (open symbols), respectively. Values consistently above the classical limit demonstrate that the quantum correlation is maintained up to this temperature. Density matrices and  RF-SET signal histograms of all Bell states and temperatures up to $\SI{1.1}{\kelvin}$ are shown in Extended Data Fig.~\ref{fig:BellStates_temperature} and Fig.~\ref{fig:Histograms_temperature}, respectively.

Figure~\ref{fig:main_fig_4}b shows the Bell state and charge readout fidelities as a function of operation temperature. The charge readout fidelity is almost unity up to $T = \SI{0.7}{\kelvin}$ and then only drops to $F_\mathrm{CRO} = \SI{99.02 \pm 0.13}{\percent}$ when the histogram peaks start to overlap significantly at $T = \SI{1.1}{\kelvin}$. Naturally, this could be improved by increasing the integration time if it were considered a limiting factor for the quantum correlation measurement. However, the decrease of the uncorrected Bell state fidelity originating from a combination of deteriorated initialization, coherence and spin-to-charge conversion is evidently the reason for approaching the classical limit. 

\section*{Bell state lifetime}

\begin{figure}[ht!]
    %\hspace{-0.6cm}
    \includegraphics[width=0.5\textwidth,angle = 0]{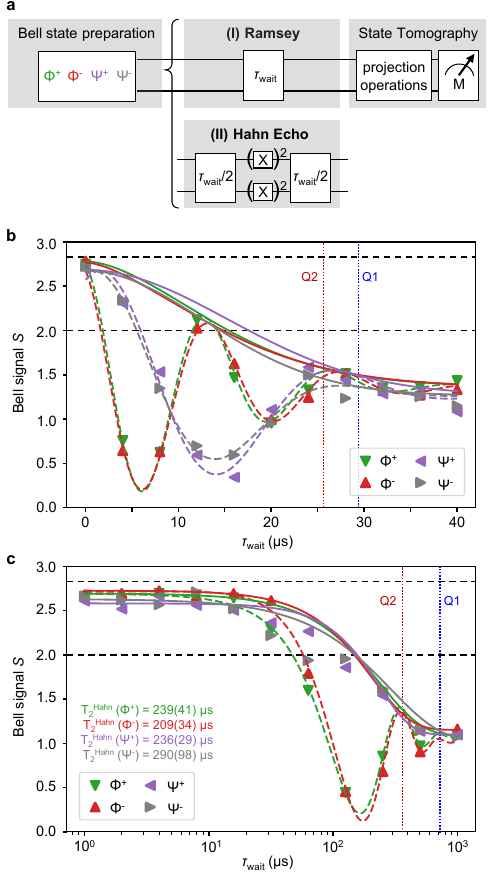}
    \caption{\textbf{Bell state lifetime.} 
    \textbf{a}, Protocol for conducting (I) Ramsey and (II) Hahn echo experiments on maximally entangled Bell states. The density matrix is measured by quantum state tomography.
    \textbf{b}, Bell signal $S$ as a function of wait time $\tau_\mathrm{wait}$ after preparation of the four maximally entangled Bell states at $T = \SI{0.1}{\kelvin}$. %Single qubit coherence time at $\SI{0.1}{\kelvin}$ is $T^\mathrm{Ramsey}_\mathrm{2,Q1} = \SI{29.37\pm 0.66}{\micro \second}$ and $T^\mathrm{Ramsey}_\mathrm{2,Q2} = \SI{26.00\pm 0.55}{\micro \second}$
    \textbf{c}, Bell signal $S$ of the four maximally entangled Bell states as a function of a total wait time $\tau_\mathrm{wait}$ being equally separated by a single, consecutive refocusing $\pi$~pulse on Q1 and Q2 at $\SI{0.1}{\kelvin}$. The Q1 and Q2 single qubit coherence times are indicated by the blue and red dashed lines, respectively. %Single qubit coherence time at $\SI{0.1}{\kelvin}$ is $T^\mathrm{Hahn}_\mathrm{2,Q1} = \SI{715\pm 16}{\micro \second}$ and $T^\mathrm{Hahn}_\mathrm{2,Q2} = \SI{350\pm 24}{\micro \second}$.
    Error bars represent the \SI{95}{\percent} confidence level.
    }
    \label{fig:main_fig_5}
\end{figure}

\noindent After having discussed how the Bell state fidelity and level of entanglement are affected by temperature, in this section we focus on the effect of idling time. This is particularly relevant for quantum information purposes when considering running an actual quantum circuit. Figure~\ref{fig:main_fig_5}a shows the protocol for Ramsey and Hahn Echo experiments on Bell states measured using quantum state tomography. 
Figure~\ref{fig:main_fig_5}b shows the Bell signal as a function of wait time after state preparation at base temperature. We observe the Bell states undergoing decoherence and staying above the classical limit ($S > 2$) for about $\SI{15}{\micro \second}$. 
Varying detunings of Q1 and Q2 from their respective Larmor frequencies result in the different oscillations frequencies. Here, $\Phi$ and $\Psi$ states are naturally grouped together due to their respective symmetric and anti-symmetric character that leads to the same accumulation of phase. We find a small correlation coefficient of $\rho = \SI{0.15 \pm 0.14}{}$ assuming a Gaussian quasi-static noise model~\cite{boter_spatial_2020}.

The lifetime of maximally entangled Bell states can be prolonged by an order of magnitude to above $\SI{100}{\micro \second}$ when applying a Hahn echo refocusing pulse (Fig.~\ref{fig:main_fig_5}c), and we expect the lifetime can be prolonged even further by higher order dynamical decoupling sequences. The oscillations in the Bell signal originate from a time-correlated nature of the IZ and ZI noise in the spin system. The decay times of the envelopes are extracted from exponential fits to the square sum of the state's Pauli projections (Extended Data Fig.~\ref{fig:PauliStates_decay}). We do not observe significant spatial correlation during Hahn echo experiments since the product of the single qubit decays $T^\mathrm{Hahn}_\mathrm{2,Q1*Q2} = \SI{235 \pm 21}{\micro \second}$ matches the Bell state lifetimes.
At higher temperatures the Bell state Ramsey and Hahn lifetimes decrease from around $\SI{20}{\micro \second}$ and $\SI{250}{\micro \second}$ to $\SI{5}{\micro \second}$ and $\SI{50}{\micro \second}$, respectively (Extended Data Fig.~\ref{fig:Coherence_temp}, Fig.~\ref{fig:BellRamsey_fitted_temp}, and Fig.~\ref{fig:BellHahn_fitted_temp}). The temporal and spatial noise correlations are unchanged (Extended Data Fig.~\ref{fig:BellRamsey_raw_temp} and Fig.~\ref{fig:BellHahn_raw_temp}).

\section*{Conclusions}
The deterministic preparation, storage and measurement of the maximally entangled quantum states that violate Bell’s inequality with $S = \SI{2.731 \pm 0.088}{}$ at $\SI{0.1}{\kelvin}$ and $\SI{2.142 \pm 0.109}{}$ at $\SI{1.1}{\kelvin}$ provides a milestone for quantum information processing with gate-defined quantum dots in silicon. Systematically reducing the error sources, carefully identified during this study, will allow us to improve operation fidelities further and bring the fundamental quantum limit~\cite{cirelson_quantum_1980} even closer. We also expect those improvements and longer dynamical decoupling sequences to further enhance the capabilities to prolong the lifetime of entanglement stored in a quantum circuit required for computation.

Evidently, incoherent dephasing errors are the dominating source of gate infidelities. In the future, we expect to increase qubit operation fidelities by improving the quality of the Si/SiO$_2$ interface and the SiO$_2$ layer as well as implementing more sophisticated, real-time phase tracking methods in the experimental setup. Additionally, the fabrication of SiMOS devices in industrial foundries~\cite{zwerver_qubits_2022,gonzalez-zalba_scaling_2021} will bring a reduction in defects, charge impurities~\cite{elsayed_low_2022,saraiva_materials_2022}, and residual ${}^{29}$Si, which will increase qubit coherence times and decrease required feedback schemes. 

GST enables us to develop error-dependent, tailored control pulse shapes to mitigate coherent errors arising from miscalibration and parameter drifts, as demonstrated by successfully implementing Hamiltonian phase corrections. Furthermore, we identified dephasing during free precession of the idling qubit as the major infidelity contribution in this study. To realize a full-scale fault-tolerant quantum computer based on silicon spin qubits, we need scalable control techniques, such as the multi-qubit SMART protocol~\cite{hansen_pulse_2021,seedhouse_quantum_2021,hansen_implementation_2022,hansen_entangling_2023}. In the SMART protocol, qubits are continuously driven by a modulated microwave field, which decouples the qubits from noise and eliminates free precession, and hence the major infidelity source. Scaling up the number of high-fidelity qubits will enable us to extend this work to quantum correlation measurements on a tripartite~\cite{coffman_distributed_2000} or multipartite~\cite{osborne_general_2006} system to violate Mermin's inequality~\cite{mermin_extreme_1990} in ever larger gate-defined quantum dot processors.

\clearpage
\newpage

%Methods – up to 3,000 additional words, may contain subsections. Methods appear online only.
\section*{Methods}
\setcounter{section}{0}

\section{Measurement setup}\label{methods:measurement_setup}
\noindent The device is measured in a K100 Kelvinox dilution refrigerator and mounted on the cold finger. Up to $T=\SI{1.1}{\kelvin}$, elevation from the base temperature is achieved by switching on and tuning the heater near the sample. Stable temperatures above $\SI{1.1}{\kelvin}$ can only be achieved by reducing the amount of He mixture in the circulation and consequently the cooling power.

An external DC magnetic field is supplied by an IPS120-10 oxford superconducting magnet. The magnetic field points along the [110] direction of the Si lattice. DC voltages are supplied with a QDevil QDAC, through DC lines with a bandwidth from 0 to $\SI{100}{\hertz}$ - $\SI{10}{\kilo \hertz}$. Dynamic voltage pulses are generated with a Quantum Machines OPX+ and combined with DC voltages via custom voltage combiners on top of the refrigerator at room temperature. The OPX+ has a sampling time of $\SI{4}{\nano\second}$. The dynamic pulse lines in the fridge have a bandwidth of 0 to $\SI{50}{\mega \hertz}$, which translates into a minimum rise time of $\SI{20}{\nano\second}$. Microwave pulses are synthesised using a Keysight PSG8267D Vector Signal Generator, with the baseband I/Q and pulse modulation signals supplied by the OPX+. The modulated signal spans from $\SI{250}{\kilo\hertz}$ to $\SI{44}{\giga\hertz}$, but is band-limited by the fridge line and a DC block.

The charge sensor comprises a single-island SET connected to a tank circuit for reflectometry measurement. The return signal is amplified by a Cosmic Microwave Technology CITFL1 LNA at the $\SI{4}{\kelvin}$ stage followed by two Mini-circuits ZFL-1000LN+ LNAs at room temperature. The Quantum Machines OPX+ generates the tones for the RF-SET, and digitises and demodulates the signals after the amplification.

\section{Bell's theorem and quantum correlation}\label{methods:quantum_correlation}
%%% Quantum correlation, State Tomography, and CHSH inequality
The quantum correlations of the spin pairs are
\begin{align}
\begin{split}
    E & = \dfrac{N_{\uparrow\uparrow}-N_{\uparrow\downarrow}-N_{\downarrow\uparrow}+N_{\downarrow\downarrow}}{N_{\uparrow\uparrow}+N_{\uparrow\downarrow}+N_{\downarrow\uparrow}+N_{\downarrow\downarrow}} \\
    & = \dfrac{N_{\uparrow\uparrow}-N_{\uparrow\downarrow}-N_{\downarrow\uparrow}+N_{\downarrow\downarrow}}{N_{\text{total}}} \\
    & = \dfrac{N_{\text{even}}-N_{\text{odd}}}{N_{\text{total}}} \\
    & = P_{\text{even}} - P_{\text{odd}} \\
    & = 2P_{\text{even}} - 1,
\end{split}
\end{align}
while assuming parity readout. The Clauser-Horne-Shimony-Holt (CHSH) inequality~\cite{clauser_proposed_1969} then becomes
\begin{align}
\begin{split}
     S & = \langle\alpha,\beta\rangle - \langle\alpha,\beta'\rangle + \langle\alpha',\beta\rangle + \langle\alpha',\beta'\rangle \\
     & = E(\alpha,\beta) - E(\alpha,\beta') + E(\alpha',\beta) + E(\alpha',\beta') \\
     & = 2(P_{\text{even}}^{\alpha,\beta} - P_{\text{even}}^{\alpha,\beta'} + P_{\text{even}}^{\alpha',\beta} + P_{\text{even}}^{\alpha',\beta'} - 1) \\
     & \geq 2,
\end{split}
\end{align}
with combinations of measurement basis $\alpha=0$, $\beta=\pi/4$, $\alpha'=-\pi/2$, and $\beta'=-\pi/4$. For the four different Bell states we have to adjust the signs/measurement axis to 
\begin{align*}
\begin{split}
     S^{\Phi+} & = E(\alpha,\beta) + E(\alpha,\beta') + E(\alpha',\beta) - E(\alpha',\beta') \\
     & = 2(P_{\text{even}}^{\alpha,\beta} + P_{\text{even}}^{\alpha,\beta'} + P_{\text{even}}^{\alpha',\beta} - P_{\text{even}}^{\alpha',\beta'} - 1),
\end{split}
\end{align*}

\begin{align*}
\begin{split}
     S^{\Phi-} & = E(\alpha,\beta) - E(\alpha,\beta') + E(\alpha',\beta) + E(\alpha',\beta') \\
     & = 2(P_{\text{even}}^{\alpha,\beta} - P_{\text{even}}^{\alpha,\beta'} + P_{\text{even}}^{\alpha',\beta} + P_{\text{even}}^{\alpha',\beta'} - 1),
\end{split}
\end{align*}

\begin{align*}
\begin{split}
     S^{\Psi+} & = -E(\alpha,\beta) - E(\alpha,\beta') - E(\alpha',\beta) + E(\alpha',\beta') \\
     & = 2(-P_{\text{even}}^{\alpha,\beta} - P_{\text{even}}^{\alpha,\beta'} - P_{\text{even}}^{\alpha',\beta} + P_{\text{even}}^{\alpha',\beta'} + 1),
\end{split}
\end{align*}

and

\begin{align*}
\begin{split}
     S^{\Psi-} & = -E(\alpha,\beta) + E(\alpha,\beta') - E(\alpha',\beta) - E(\alpha',\beta') \\
     & = 2(-P_{\text{even}}^{\alpha,\beta} + P_{\text{even}}^{\alpha,\beta'} - P_{\text{even}}^{\alpha',\beta} - P_{\text{even}}^{\alpha',\beta'} + 1).
\end{split}
\end{align*}
We measure the even parity of the spin states after the physical rotation to the $(\alpha,\beta)$, $(\alpha,\beta')$, $(\alpha',\beta)$, and $(\alpha',\beta')$ bases. 

%%% section{State Tomography and CHSH Inequality}
An alternative approach is to measure the Bell states' density matrix $\rho$ with quantum state tomography and apply the respective rotation analytically. For that, we apply the rotation matrix
\begin{align}
    R(\varphi) = 
    \begin{bmatrix}
        \cos{\varphi/2} & -i\sin{\varphi/2}\\
         -i\sin{\varphi/2}& \cos{\varphi/2}
    \end{bmatrix}
\end{align}
with $\alpha=0$, $\beta=\pi/4$, $\alpha'=-\pi/2$, and $\beta'=-\pi/4$ to calculate the even parity 
\begin{align}
\begin{split}
    P_{\text{even}}^{\alpha\beta} & = P_{\downarrow\downarrow}^{\alpha\beta} + P_{\uparrow\uparrow}^{\alpha\beta} \\
    & = \left[R(\alpha) \otimes R(\beta) \times \rho \times \overline{R(\alpha)\otimes R(\beta)}\right]_{00} \\
    & + \left[R(\alpha) \otimes R(\beta) \times \rho \times \overline{R(\alpha)\otimes R(\beta)}\right]_{33}.
\end{split}
\end{align}

\section{Error taxonomy with pyGSTi}\label{methods:error_taxonomy}
\noindent When examining the fidelity results, we are also interested in understanding the dominant error sources behind the $\mathrm{XI}$, $\mathrm{IX}$, and $\mathrm{DCZ}$ gate infidelity. To categorise the gate errors, we use gate set tomography for decomposing errors implemented in the pyGSTi package~\cite{nielsen_pygstiopygsti_2022, blume-kohout_taxonomy_2022}.

\section*{Acknowledgements}
\noindent We acknowledge technical support from Alexandra Dickie and Rodrigo Ormeno Cortes. We acknowledge technical discussions on GST with Corey Ostrove, Kenneth Rudinger, and Robin Blume-Kohout. We acknowledge support from the Australian Research Council (FL190100167 and CE170100012), the U.S. Army Research Office (W911NF-23-10092), the U.S. Air Force Office of Scientific Research (FA2386-22-1-4070), and the NSW Node of the Australian National Fabrication Facility. The views and conclusions contained in this document are those of the authors and should not be interpreted as representing the official policies, either expressed or implied, of the Army Research Office, the U.S. Air Force or the U.S. Government. The U.S. Government is authorised to reproduce and distribute reprints for Government purposes notwithstanding any copyright notation herein. P.S., M.K.F., S.S., R.Y.S., J.Y.H., and C.J. acknowledge support from Sydney Quantum Academy. P.S. acknowledges support from the Baxter Charitable Foundation.

\section*{Author contributions}
\noindent P.S., T.T, A.S., C.H.Y., A.S.D., and A.L. designed the experiments. P.S. performed the experiments under A.S., C.H.Y., A.S.D., and A.L.'s supervision. W.H.L. and F.E.H. fabricated the device under A.S.D.'s supervision on enriched $^{28}$Si wafers supplied by K.M.I.. S.S. designed the RF-SET setup. N.D.S., S.S., E.V., and A.L. contributed to the experimental hardware setup. N.D.S. and S.S. contributed to the experimental software setup. J.Y.H., A.S., and C.H.Y. designed the heralded initialisation protocol. M.K.F. assisted with the two-qubit sequence generation. P.S. performed the subsequent error generator analysis with pyGSTi under M.K.F.'s supervision. T.T., W.H.L., N.D.S., M.K.F., S.S., R.Y.S., C.J., C.C.E., A.M., A.S., C.H.Y., A.S.D., and A.L. contributed to the discussion, interpretation and presentation of the results. P.S, T.T., A.M., A.S., C.H.Y., A.S.D., and A.L. wrote the manuscript, with input from all co-authors.

\section*{Corresponding authors}
\noindent Correspondence to P.S., A.S.D., or A.L..

\section*{Competing interests}
\noindent A.S.D. is the CEO and a director of Diraq Pty Ltd. T.T., W.H.L., N.D.S., E.V., F.E.H., C.C.E., A.S., C.H.Y., A.S.D., and A.L. declare equity interest in Diraq Pty Ltd..

\section*{Data availability}
\noindent All data of this study will be made available in an online repository.

\section*{Code availability}
\noindent The analysis codes that support the findings of the study are available from the 
corresponding authors on reasonable request.

\bibliographystyle{naturemag}
\bibliography{main.bib}

%%%Extended data figures – up to 10 items (figures and/or tables), appearing online only.
\clearpage
\newpage
\onecolumngrid
\vfill

%\section*{Extended data}
\setcounter{figure}{0}
\setcounter{table}{0}
\captionsetup[figure]{name={\bf{Extended Data Fig.}},labelsep=line,justification=centerlast,font=small}

\begin{figure*}[ht!]
    %\hspace{-0.6cm}
    \includegraphics[angle = 0]{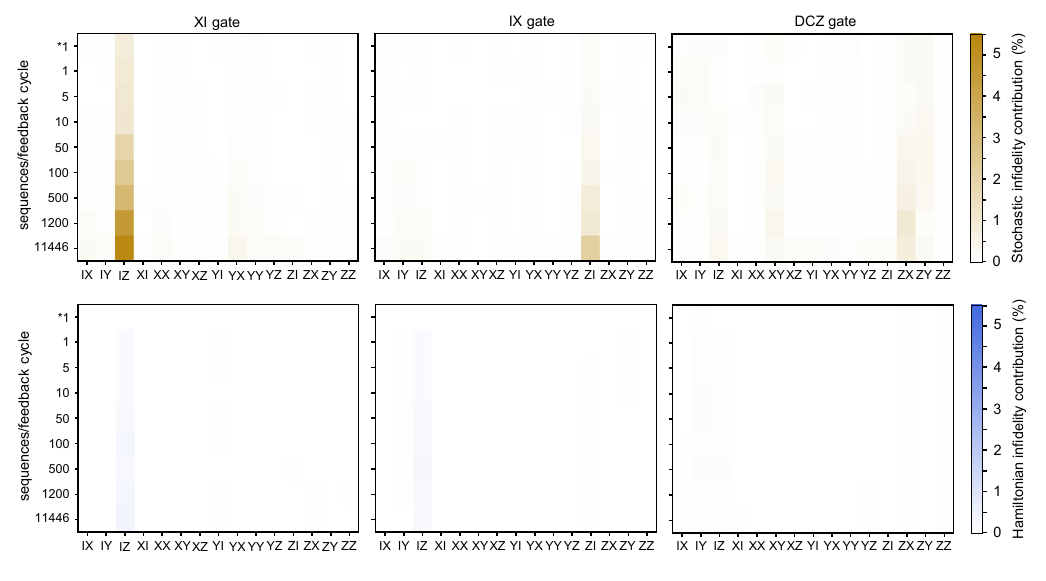}
    \caption{\textbf{Breakdown of GST error channels.}
    Infidelity contributions of individual error channels using pyGSTi on Larmor frequency feedback rate measurement series including the additionally phase corrected experiment (*) with fidelities of full 2-qubit gate set above \SI{99}{\percent}.} 
    \label{fig:LarmorFeedback_GST_breakdown}
\end{figure*}

\begin{figure*}[ht!]
    %\hspace{-0.6cm}
    \includegraphics[angle = 0]{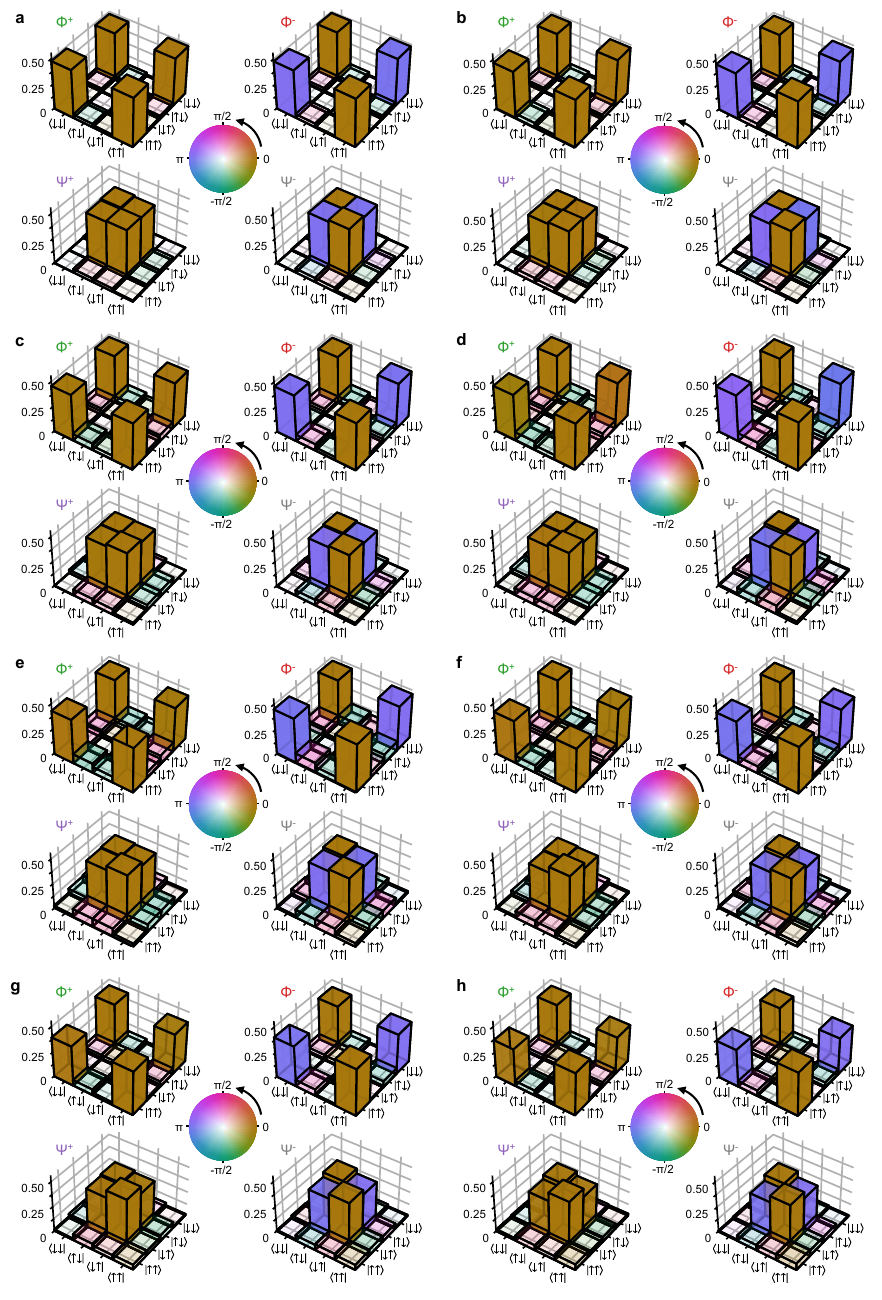}
    \caption{\textbf{Bell state tomography as a function of temperature.} 
    Density matrices of all four Bell states for temperatures \textbf{a}, \SI{0.1}{\kelvin}, \textbf{b}, \SI{0.2}{\kelvin}, \textbf{c}, \SI{0.3}{\kelvin}, \textbf{d}, \SI{0.5}{\kelvin}, \textbf{e}, \SI{0.7}{\kelvin}, \textbf{f}, \SI{0.85}{\kelvin}, \textbf{g}, \SI{1.0}{\kelvin}, and \textbf{h}, \SI{1.1}{\kelvin} extracted from quantum state tomography.
    }
    \label{fig:BellStates_temperature}
\end{figure*}

\begin{figure*}[ht!]
    %\hspace{-0.6cm}
    \includegraphics[angle = 0]{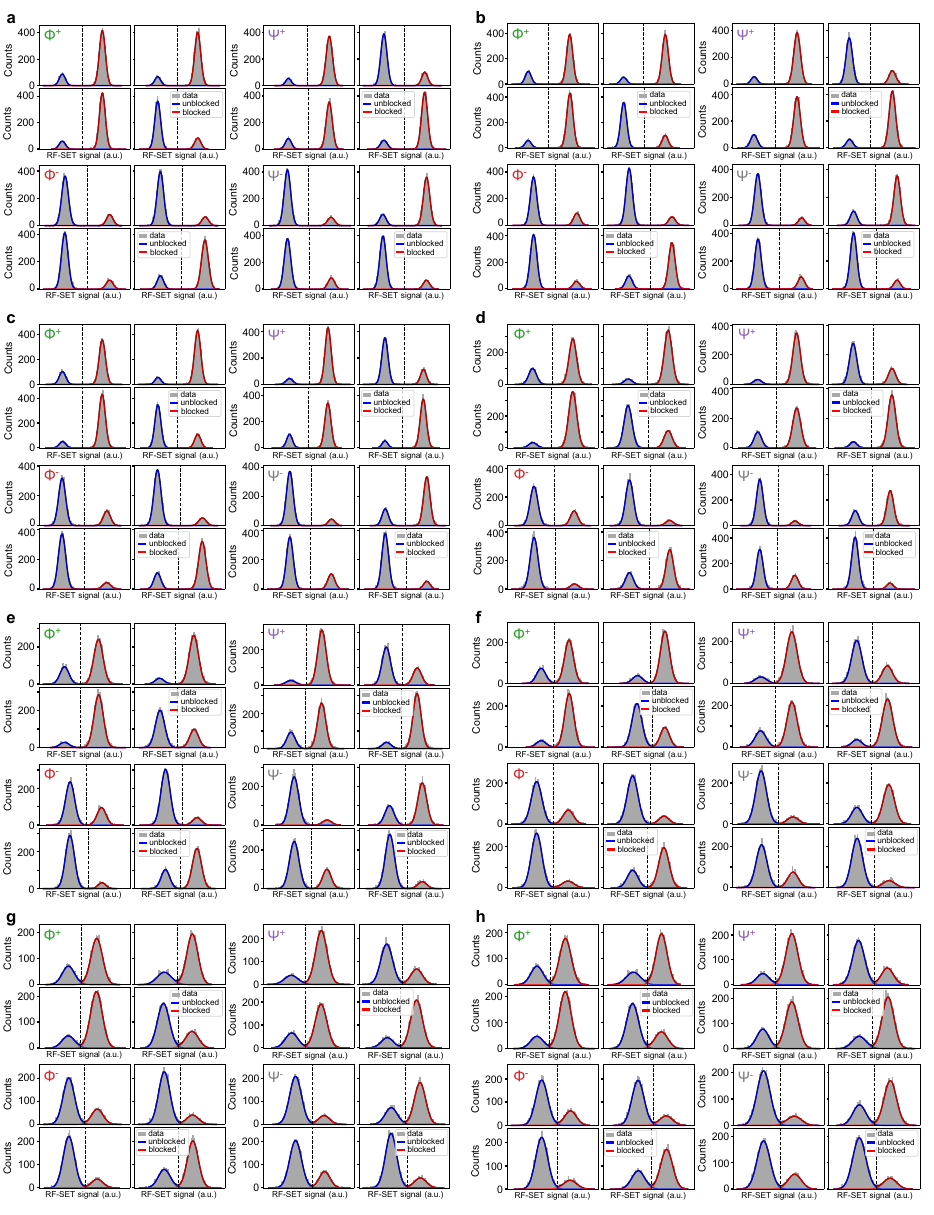}
    \caption{\textbf{RF-SET signal histograms as a function of temperature.} 
     RF-SET signal histograms of all four Bell states for temperatures \textbf{a}, \SI{0.1}{\kelvin}, \textbf{b}, \SI{0.2}{\kelvin}, \textbf{c}, \SI{0.3}{\kelvin}, \textbf{d}, \SI{0.5}{\kelvin}, \textbf{e}, \SI{0.7}{\kelvin}, \textbf{f}, \SI{0.85}{\kelvin}, \textbf{g}, \SI{1.0}{\kelvin}, and \textbf{h}, \SI{1.1}{\kelvin} obtained during the direct parity Bell experiment.
    }
    \label{fig:Histograms_temperature}
\end{figure*}

\begin{figure*}[ht!]
    %\hspace{-0.6cm}
    \includegraphics[angle = 0]{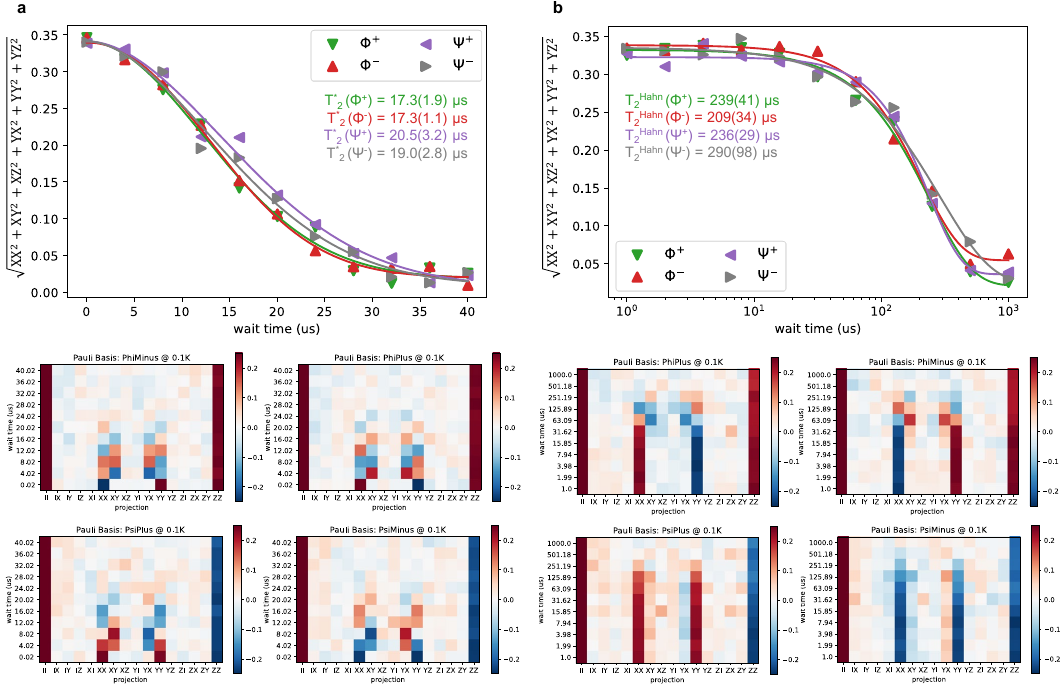}
    \caption{\textbf{Pauli projections of Bell state lifetime experiments.} 
    Pauli projections of all four Bell states as a function of wait time for \textbf{a}, Ramsey and \textbf{b}, Hahn experiment. The lower panels show the individual Pauli projections. In the upper panels we fit the squared sum of the contributing Paul projections to an exponential decay to extract all four Bell state lifetimes.
    Error bars represent the \SI{95}{\percent} confidence level.
    }
    \label{fig:PauliStates_decay}
\end{figure*}

\begin{figure*}[ht!]
    %\hspace{-0.6cm}
    \includegraphics[angle = 0]{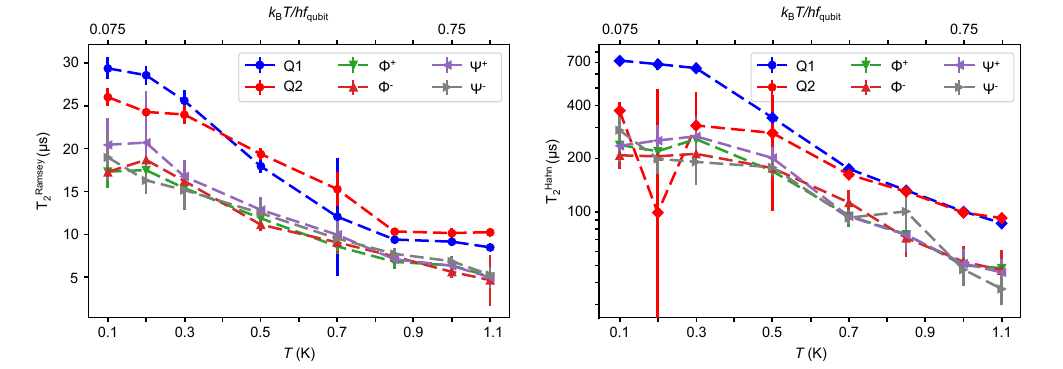}
    \caption{\textbf{Coherence times as a function of temperature.} 
    \textbf{a}, Ramsey and \textbf{b}, Hahn echo coherence times of the single qubits and all four Bell states as a function of temperature.
    Error bars represent the \SI{95}{\percent} confidence level.
    }
    \label{fig:Coherence_temp}
\end{figure*}

\begin{figure*}[ht!]
    %\hspace{-0.6cm}
    \includegraphics[angle = 0]{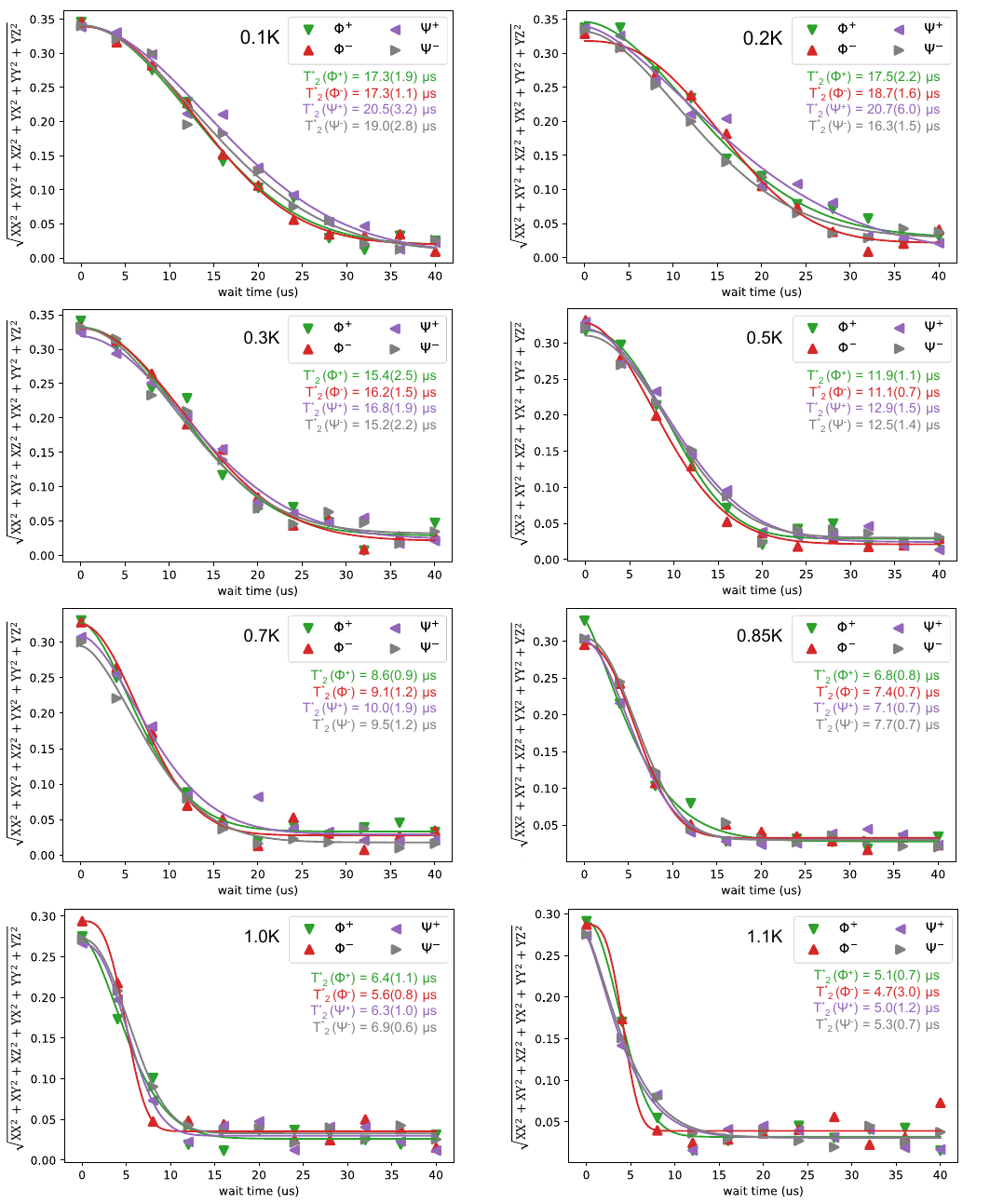}
    \caption{\textbf{Ramsey Pauli projections as a function of temperature -- squared sum.} The solid lines are exponential decays fitted to the data to extract the coherence times. 
    Error bars represent the \SI{95}{\percent} confidence level.
    }
    \label{fig:BellRamsey_fitted_temp}
\end{figure*}

\begin{figure*}[ht!]
    %\hspace{-0.6cm}
    \includegraphics[angle = 0]{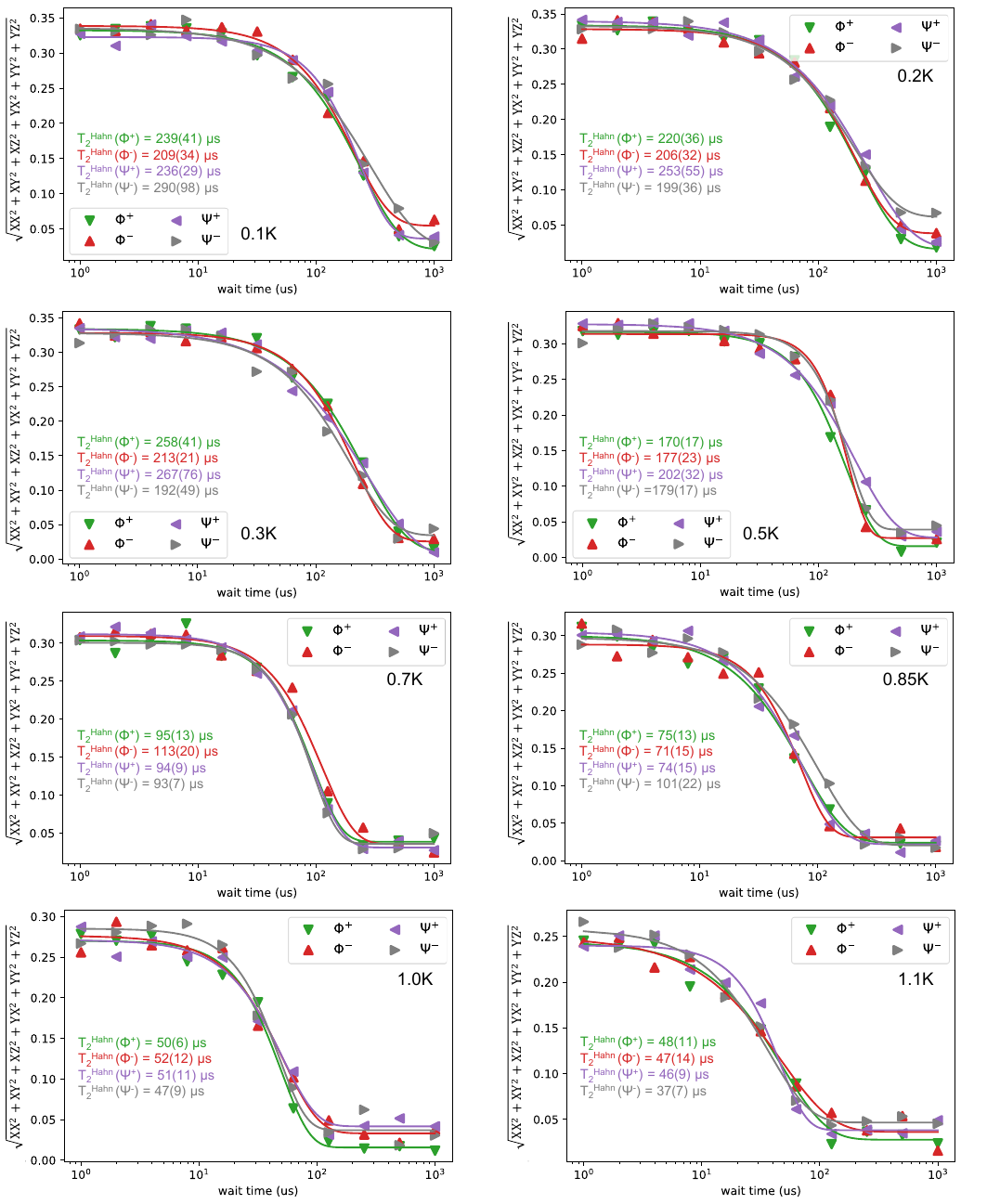}
    \caption{\textbf{Hahn echo Pauli projections as a function of temperature -- squared sum.} The solid lines are exponential decays fitted to the data to extract the coherence times.
    Error bars represent the \SI{95}{\percent} confidence level.
    }
    \label{fig:BellHahn_fitted_temp}
\end{figure*}

\begin{figure*}[ht!]
    %\hspace{-0.6cm}
    \includegraphics[angle = 0]{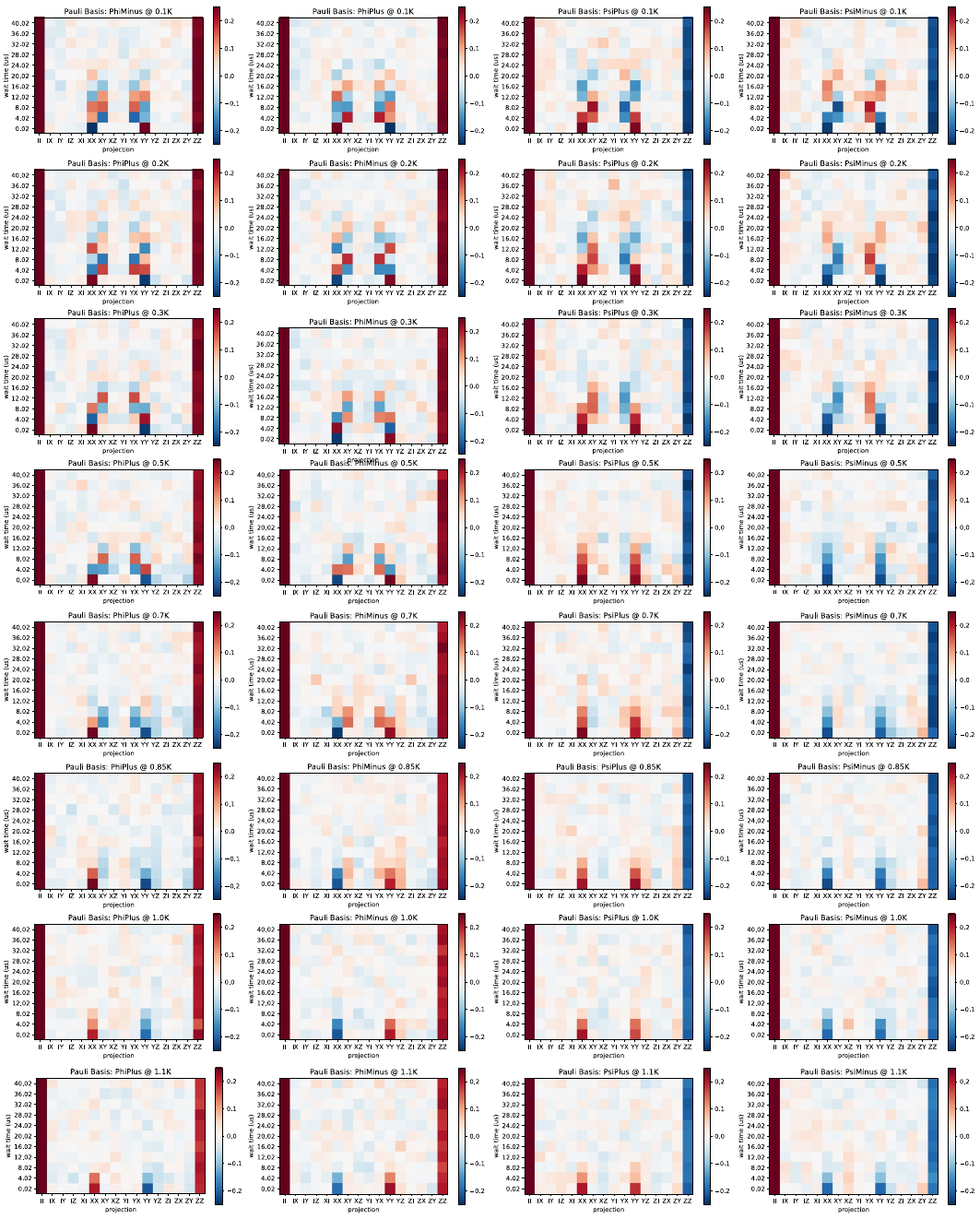}
    \caption{\textbf{Ramsey Pauli projections as a function of temperature -- individual projections.} 
    }
    \label{fig:BellRamsey_raw_temp}
\end{figure*}

\begin{figure*}[ht!]
    %\hspace{-0.6cm}
    \includegraphics[angle = 0]{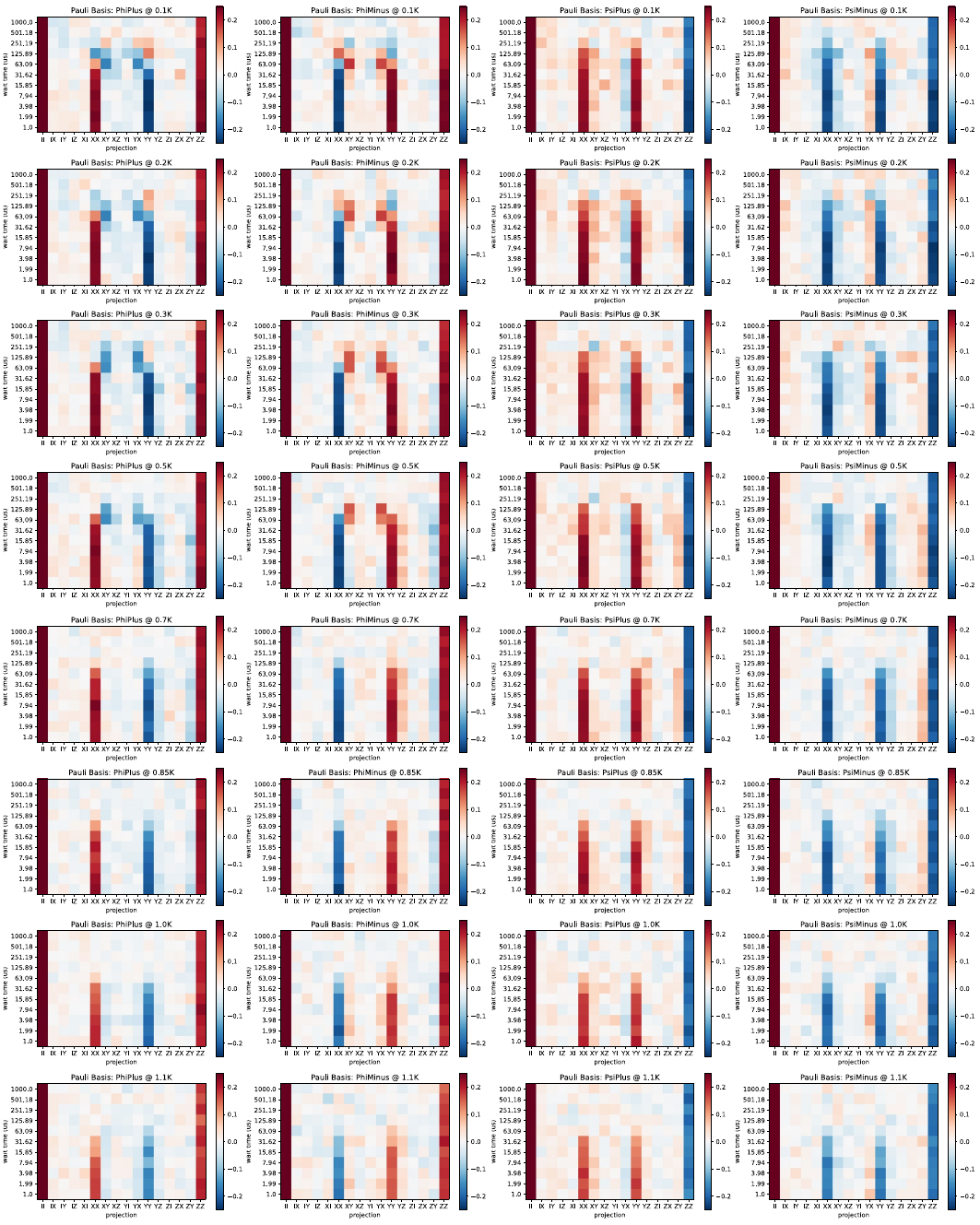}
    \caption{\textbf{Hahn echo Pauli projections as a function of temperature -- individual projections.} 
    }
    \label{fig:BellHahn_raw_temp}
\end{figure*}

\setcounter{figure}{0}
%\captionsetup[figure]{name={\bf{Supplementary Fig.}},labelsep=line,justification=centerlast,font=small}

\end{document}